\documentclass[letterpaper, 10 pt, conference]{ieeeconf}  
\usepackage{indentfirst}        
\usepackage{graphicx}
\usepackage{url}
\usepackage{hyperref}
\hypersetup{
    colorlinks=true,
    linkcolor=cyan,
    filecolor=magenta,      
    urlcolor=cyan,
}

\IEEEoverridecommandlockouts                              
\title{\LARGE \bf
VeriSign: A Secure Contract Consensus Platform on the Blockchain with Amendment Functionality
}

\author{
  \textbf{Mustafa Bal}, \textbf{Rangel Milushev}, \textbf{Kaan Armagan}\\
  \texttt{\{mbal,rmilushev,armagan\}@college.harvard.edu} \\
  Harvard University\\
  December 10, 2018
}

\begin{document}

\maketitle
\thispagestyle{empty}
\pagestyle{empty}

\begin{abstract}

While electronic signatures are widespread, there currently exists no viable signing solutions that can track amendments. We proposed VeriSign, a secure contract consensus platform where amendments to contracts can be tracked in a decentralized medium. We demonstrate a user-facing app where signatories can vote on original contracts and amendments, and incorporate a Blockchain where we store the transaction history of original contracts and amendments. This platform has possible applications in tracking the history of legislation, and amendments to legislation.

\end{abstract}

\section{Introduction} 

\subsection{Electronic Signatures}

Electronic signatures have become a popular and effective means of allowing people to sign and send documents without needing to have all the signatories in the same room, and without having to print out physical copies or having to pay postal services to send a document to a signatory. These signatures are most certainly legally binding even if they do not exist in the physical world; there are many cases in which judges have ruled in favor of the veracity of eSignatures \cite{publaw}. Whenever there's consent and intent, which is almost always the case, eSignatures are legally binding. In 2000, the United States passed the ESIGN Act, making e-signatures legal for virtually all uses. In the European Union the Electronic Identification and Trust Services Regulation (eIDAS) took effect in July 2016. Similar laws giving legal validity to electronic signatures have been passed and are in effect in many countries around the world \cite{globalsign}. 

\subsection{Existing Solutions}

There currently are many solutions both on and off the Blockchain that enable users to sign, transfer, notarize and even audit legal documents \cite{signatura}. On the side of non-Blockchain methods, there are very simple software solutions such as Adobe Sign that simply allow users to put their signatures on documents \cite{adobesigncap}. There are also more complex and functional systems such as HelloSign\cite{cHelloSign} that allow users to send and sign documents electronically. One of the main problems with solutions such as HelloSign is that they store the user's documents on a centralized system, creating concerns over users' privacy as well as the safety and integrity of their documents. If a signed document were to be altered in any way without the unanimous consent of all signatories, the document automatically loses its legal validity. On the other hand, there exist Blockchain solutions such as BlockSign \cite{blocksign} and Signatura \cite{signatura} that address this problem by utilizing distributed ledgers, smart contracts and data immutability to address the problems of privacy, safety and integrity of users' documents. 

\subsection{Our Solution}

There are many unique and useful features that these solutions offer; however, none of these existing solutions have been able to implement a feature that allows users to amend documents while also maintaining the legal integrity of the document. Currently, to be able to make an amendment digitally, a user would need to create a new version of the document they would like to amend. This new version of the document is not linked in any way to the original document; thus the user will need to track down all the original signatories to get them to approve this new version. Our goal is to streamline this experience of amending documents digitally by creating a Blockchain solution that enables users to sign, send and amend documents by obtaining the approval of all the original signatories of the first version of the document. This solution would expedite the bureaucratic processes of academic institutions, enterprises and possibly even countries as these entities go through a natural regulatory evolution over time which involves making amendments in already existing laws, by-laws, rules and beyond. Making these amendments via our solution would add an additional layer of transparency to the process of regulatory change as well as enabling more effective and safer record-keeping of the history of legal documents.

\section{Proposed approach}

\subsection{Integrating a consensus platform with Blockchain}
To achieve a platform where users are able to approve original contracts and amendments to a given document in a decentralized and secure manner, we utilize an end-to-end encrypted consensus platform and a Blockchain solution in parallel. 

The consensus platform is where users of VeriSign who are signatories of a given contract approve or reject texts of contracts. This platform stores the complete .txt version of the contract in question, which includes sections such as the date of the agreement, main body, signatories, and approved amendments. The consensus platform also stores the unique SHA-256 hash value of the unique contract in .txt formatting. This allows all users to check the integrity of the contract they sign: even a small change in a given document results in a completely different SHA-256 hash value, thus all users are able to make sure the contract they are signing is the exact one they intend to sign, and that all users are signing the same equal contract. Signatories are able to issue their votes, consisting of "yes" and "no" votes, on the proposed document. All users are able to see how others have voted.


The Blockchain platform is where the history of the contracts are stored. An original or an amendment contract has a "VeriOwner" entity who is the actual owner of the given contract. This VeriOwner is generated by VeriSign when all signatories of a given contract give their approval for that contract. Each VeriOwner entity has its own public and private keys, which are used when a transaction is added to the Blockchain. The genesis block on the Blockchain is manually added so that the next and following blocks, which will contain the SHA-256 hashes of the contract as well as the owner of that contract, can be added to the Blockchain. 

\subsection{Issuing the original contract}

\begin{figure}[thpb]
      \centering
      \includegraphics[scale=0.50]{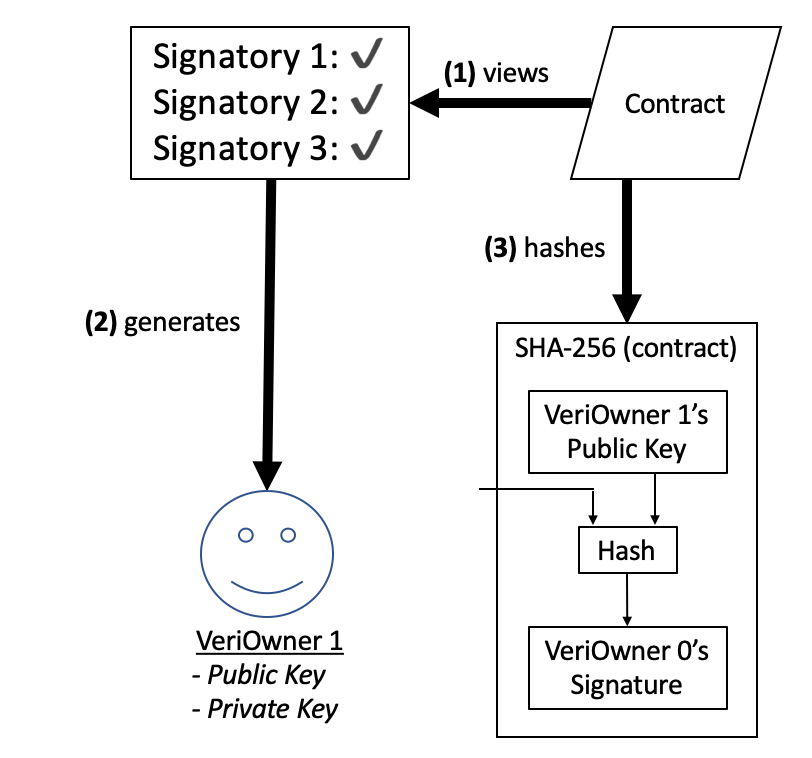}
      \caption{Creating the block for a contract. In this example, a block is generated for an original document.}
      \label{figurelabel}
   \end{figure}

To create an original contract, a VeriSign consensus group is first formed on the VeriSign platform. All signatories of this contract are added to the same consensus group. The text of the proposed contract is displayed where all signatories can read and evaluate. The signatories then issue their votes on the proposed document. If at least one voter issues a "no" vote, that proposition fails, and a new proposition begins after necessary changes are made to the contract text. Once all signatories give "yes" as their votes, a VeriOwner is generated by the VeriSign consensus group. This VeriOwner becomes the actual owner of the document, and its ownership is displayed to the users by recording the contract on the Blockchain with the new VeriOwner. 

To document this original contract on the Blockchain, a block is first created that contains the SHA-256 hash of the contract's text. This block also contains the hash of the previous block and the current contract owner's public key as well as the signature of the previous block's VeriSign owner.

\subsection{Issuing amendments}

To create an amendment of an already existing original VeriSign contract, the amendment is proposed in the VeriSign consensus group that was first created for the original VeriSign contract. Again, all signatories need to vote "yes" to approve the amended contract. Once this condition is satisfied, a new VeriOwner entity is generated by the VeriSign consensus platform with its own public and private keys. This new VeriOwner then becomes the owner of the amended contract. 

The amended contract's SHA-256 hash value, along with the newly generated VeriOwner is then added to the Blockchain. This block also contains the has of the previous block, the current contract owner's public key, and the signature of the previous block's VeriOwner.

\begin{figure}[thpb]
      \centering
      \includegraphics[scale=0.52]{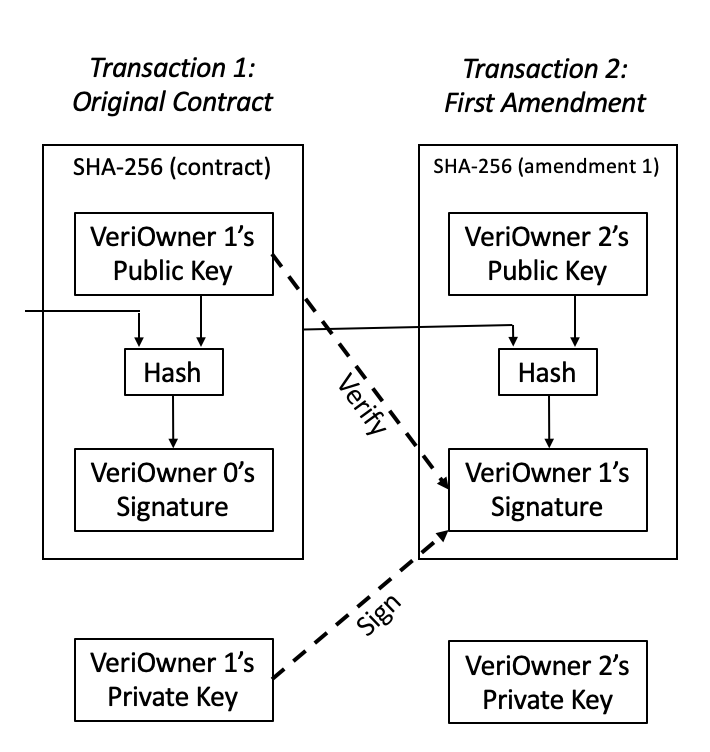}
      \caption{Issuing an amendment for an original contract.}
      \label{figurelabel}
   \end{figure}

\subsection{Advantages of the VeriSign consensus platform}

The VeriSign consensus platform offers multiple advantages that previous forays in the e-Signatures do not posses. VeriSign provides an electronic medium to track approved amendments in a decentralized manner. As each approved contract's history is published on the blockchain, it is very easy for users to confirm the exact amendments to given contracts have been made. 

In addition, all signatories can guarantee that the contract they are signing is exactly the same as the one that other signatories to the contract are signing, thereby eliminating information inconsistency issues and thereby assuring its validity. If a contract even has a single bit modified, the SHA-256 hash codes of the two documents in question will be different, and so the modified document will be seen as false. Thus, the VeriSign consensus platform guarantees the security, integrity and the validity of all original and amendment contracts.

\subsection{VeriSign Success Criteria}

The success of the VeriSign Signing and Consensus Platform is measured by multiple benchmarks. First, the platform should be able to track the history of the original and amended contracts. Secondly, the platform should be able to guarantee that all users of the system are signing the same contracts as each other. Thirdly, the approval process of both original and amended contracts should be secure in which only the signatories of these contracts are able to vote on their approvals, and that these contracts can be approved if and only if all signatories unanimously vote affirmatively on the proposed contracts.

\section{Work performed}
Given the time constraint and limited scope of our work, we have not been able to completely create a blockchain application that manages to fulfill the approach described above. What we have developed is a prototype of how our model would function.
\subsection{Blocks \& Miners}
In our implementation, each block has the following components: an id number, the id number of the miner that generated it, the time-stamp when the block was generated, the data stored, the hash of the previous block, and the hash that the miner generated for the current block. The data that the miners store in the blocks is the signatories of the contract and the contract itself. For a block to appear on our simulated blockchain a miner has to generate a block with an appropriate hash (in our case, we want six leading zeroes which is our oversimplified version of proof of work; this is done in the \textit{hash\_block} function). Once such a hash is generated, every miner is to check whether the has generated was actually correspondent to the truth, so they will hash the exact same parameters to verify that the hash is valid. We do this through the \textit{verify\_hash} function. If 51\% or more of the miners verify the block as valid, it is added to the chain.
\subsection{Signatories}
Each signatory is a party that would be bound by the legal agreement they would be signing. They provide us with the text they will be signing and we obtain the hash of that text, which we store under their unique id. We store whether they have signed the document as a Boolean value, which is false by default. 
\subsection{Add a document only after a consensus}
The consensus function is straightforward and intuitive. Before we hash the information into the next block, every miner checks to verify that all the signatories have submitted the hash of the same document, i.e. nobody is signing a version of the contract that is different. The second insurance that the \textit{consensus} function gives us is that 51\% of the miners or more have to say that the information is valid, and that all the required signatories have signed the document or its amendment.

\section{Results}We have implemented all of the above-discussed functions and simulated them in Python. We uploaded our repository to this \href{https://github.com/rangelak/BlockchainSimulationVS}{link}.

Through our simulations have seen firsthand the benefits that the decentralized consensus algorithm provides. Running the simulator in a noisy environment, where some of the miners are untruthful, the lack of a single point of failure provides us with a robust way of tackling attacks and hackers. As long as more than 51\% of the miner nodes are truthful, the integrity of the chain would not be compromised. Running our simulations with 5 miners we noticed that even at lower levels of noise, i.e. around $10$\%, the chain is at risk of failing given a truthful request. As we introduce more miners into the system, the problem is mitigated and the system would sustain even at higher levels of noise.
\section{Conclusions and Future Work}

We have achieved the goal of building the theoretical framework for the functionality of the amendment feature via the use of Blockchain and we have achieved the goal of building a simulation of how the system would function. We have not yet completed the implementation of the actual signing and amendment features via the use of Blockchain. If we were given another semester to work on this, we would learn how to code with Solidity to be able to program the smart contracts which would govern this system.

A lesson we have learned from our work on this project is that Blockchain can be a very powerful tool for preserving the integrity of legal documents and their amendments. By using this application, users will no longer need to trust a centralized third party to uphold the integrity, privacy, and security of legal documents. The use cases for this application could extend beyond the cases discussed in this paper; the same method could be applied to legislative processes. Amendments to countries' laws and regulations happen very frequently; this change is sometimes so rapid that citizens are not able to keep track of everything being changed \cite{compconst}. This is a problem because of two reasons. The first reason is that in a corrupt legislative body changes can be made to laws and regulations quicker than the other government agencies and civilians can keep track of them, thus undermining democracy. The second problem is that because it is so hard to keep track of legislative changes, citizens don't even try, leading to a population that is uninformed. An app like VeriSign that legislative bodies would be required to use could solve both these problems by providing additional transparency and integrity to legislative changes.




\section{Acknowledgment}

This project would not have been possible without the help of Professor Kung and Marcus Comiter. Along the way they have graciously provided us their time and patience. We owe  them our thanks for their help and support.


\end{document}